\documentstyle[preprint,eqsecnum,aps]{revtex}

\begin{document}
\input epsf
\draft

\title{\Large \bf Dynamic Renormalization Group Study of the $\phi^4$ Model
with Colored Noise
}
\author{\large
J. Garc\'{\i}a-Ojalvo$^{a,b}$,
J.M. Sancho$^{b}$ and H. Guo$^{c}$}
\address{\mbox{ } \\
(a) Departament de F\'{\i}sica i Enginyeria
Nuclear,\\ Escola T\`{e}cnica Superior d'Enginyers Industrials
de Terrassa\\ Universitat Polit\`{e}cnica de Catalunya,
Colom 11, E-08222 Terrassa, Spain.\\
\mbox{ } \\(b) Departament d'Estructura i Constituents de la
Mat\`{e}ria, Facultat de F\'{\i}sica
\\ Universitat de Barcelona,
Diagonal 647, E-08028 Barcelona, Spain.\\
\mbox{ } \\(c) Department of Physics, McGill University\\
Rutherford Building, 3600 University Street\\
Montr\'eal, Qu\'ebec, Canada H3A 2T8.}
\maketitle
\begin{abstract}
The non-conserved $\phi^4$ model defined by a Langevin equation with external
non-white noise is studied by means of the Dynamic
Renormalization Group.
The correlation time of the noise changes the critical
point location but does not affect the critical exponents up to
order $\varepsilon^2$. The same effect is obtained when the correlation
length of the noise is considered.
These results are shown to be in agreement with previous numerical
simulations.
\end{abstract}

\narrowtext

\section{Introduction}
\label{sec:introduction}

Langevin equations in spatially-extended systems have been
extensively used in the studies of equilibrium and nonequilibrium
phenomena where fluctuations are important. These
equations are stochastic partial differential equations in which
the fluctuations are introduced through a noise term, whose
stochastic properties are chosen according to the physical situation.
Fluctuations of internal (thermal) origin
represent microscopic degrees of freedom, and since they evolve
in spatial and temporal scales much shorter than those of the gross
variables of the system, they are assumed to be uncorrelated
(delta correlated) in space and time. In this case they are
modelled by a gaussian-and-white noise process in space and time.
On the other hand, if the origin of the fluctuations is external to
the system, the possibility of a noise with some
spatial or temporal structure has to be considered.

Langevin equations have been used in many different scenarios, one of
which has been for many years the study of Dynamical Critical Phenomena
\cite{ma76,hohenberg77}. The methodology used there is a suitable
extension of the Renormalization Group techniques developed for the
study of static critical phenomena \cite{wilson74,fisher83}. The
technique has also been generalized to study nonequilibrium systems
presenting some kind of scale invariance or criticality
\cite{forster77,medina89}.

In this paper we will follow this methodology as it was implemented
in Refs. \cite{forster77,medina89} to study the nonequilibrium critical
properties of a Langevin-equation model with a non-white
(colored) noise.

Among the different models that can be considered
we have chosen the simplest
one, in which the order parameter is not conserved. This model is
useful for describing order-disorder transitions. In this
non-conserved case the equation of motion of the field variable is
\begin{equation}
\frac{\partial \psi({\bf x},t)}{\partial t} = - \frac{\delta F[\psi]}
{\delta
\psi({\bf x},t)} +\xi({\bf x},t),
\label{eq:tdglm}
\end{equation}
where ${\bf x}$ is the position in a d-dimensional space, $F[\psi]$
is the Ginzburg-Landau free-energy functional
\begin{equation}
\label{eq:glf}
F[\psi]=\int d^dx \left\{\frac{1}{2}r\psi^2+\frac{1}{2}\mu\mid\nabla
\psi\mid^2 + \frac{1}{4}u\psi^4\right\}.
\end{equation}
and the correlation of the noise is
\begin{equation}
\label{eq:white}
\left<\xi({\bf x},t)\;\xi({\bf x'},t')\right>=2D\;
\delta({\bf x}-{\bf x'})\;\delta(t-t'),
\end{equation}
where $D=k_B T$ to ensure the correct equilibrium steady state.
This is known as {\em
model A} in the notation of Ref. \cite{hohenberg77}. This
model comes from a coarse-graining procedure
applied to a spin model \cite{langer71}. Its behavior can
be easily understood by means of a mean-field analysis
of the potential given by the Ginzburg-Landau free energy.
When $r$ is positive this potential has a minimum at zero
field, which corresponds to a coarse-grained field equal to 0
(i.e. the spins are randomly up or down). A negative value of
$r$ leads to a potential with two nonzero symmetrical minima
(the spins have a preferred direction, either up or down). The
role of noise in this scheme
is a disordering one: when $r$ is negative and the intensity
of the noise increases, the potential barrier between the two
minima is more easily surpassed by the spins. Thus, although
the potential shape does not change, the coarse-grained field
will eventually become zero (for a high enough value of the
intensity of the noise): the spins will be disordered because
their thermal energy (i.e. the noise intensity) is so high
that they do not see the potential barrier.

Our aim in this paper is to study the model
(\ref{eq:tdglm}-\ref{eq:glf}) in the special situation characterized
by the fact that the noise term
$\xi({\bf x},t)$ is non-delta correlated.
In this case there is no fluctuation-dissipation relation
and the steady state is not the equilibrium one.
Hence we are facing with a pure non-equilibrium problem.
In a first approach to this problem we will assume that the
correlation is of the Ornstein-Uhlenbeck type:
\begin{equation}
< \xi({\bf x},t)\xi({\bf x}',t')> = \frac{D}{\tau}\;e^{-\frac{\mid
t-t' \mid}{\tau}} \delta^d({\bf x}-{\bf x}')
\label{eq:oun}
\end{equation}
This is a suitable way of modelling the noise in the case when
it is external.
$\tau$ is its {\em correlation time}.
It can be seen that in the limit $\tau\rightarrow 0$
the noise becomes white and the equilibrium situation is recovered
provided $D=k_B T$.
Na\"{\i}vely, one might say that an increasing value of $\tau$
makes the noise "softer", taking it farther from the white
character (given by $\tau\rightarrow 0$) and reducing its
effective intensity. This effect has indeed been found
numerically \cite{ojalvo92a} and approximate techniques
borrowed from stochastic theory have been developed in order
to understand this phenomenon \cite{ojalvo94a}. In this paper
we perform dynamic Renormalization-Group (DRG) calculations on
this model in order to study the effects of the correlation
time of the noise in the non-equilibrium critical behavior of the system.

This paper is organized as follows: in next Section we present a
preliminary scaling analysis to obtain some of the critical
characteristics of our model. In Sect. \ref{sec:drg-an} the
Dynamic Renormalization Group results are presented.
In Sect. \ref{sec:space} we summarize the results corresponding
to a noise colored in both space and time.
The comparison between our theoretical results and those obtained
from simulations is given in Sect. \ref{sec:comp}. Technical
details of the DRG calculations are presented in the Appendix.

\section{Preliminary analysis}
\label{sec:trivial-exp}
A glimpse of the relevance of the different terms in 
(\ref{eq:glf}) at the critical point can be obtained by
studying the effect of a scale transformation on the dynamical
equation (\ref{eq:tdglm}). Let this scale transformation be defined by
\begin{equation}
\label{eq:scales}
x=b\;\tilde{x};\;\;\;\;\; t=b^z\;\tilde{t}; 
\;\;\;\;\;\psi=b^{-a}\;\tilde{\psi},
\end{equation}
with $b>1$, which means that one is focusing on long distances
(characteristic of the critical point).
$a$ is related to the equilibrium exponent $\eta$ via $a=\frac{1}{2}
(d-2+\eta)$ \cite{hohenberg77}.
Introducing these changes into Eq. (\ref{eq:tdglm}) and making
use of the expression of the Ginzburg-Landau functional
(\ref{eq:glf}) one can see that the dynamical equation becomes
\begin{equation}
\label{eq:rescaled}
\frac{\partial\tilde{\psi}}{\partial \tilde{t}} =
- r\;b^z\;\tilde{\psi} + \mu\;b^{z-2}\;\nabla^2
\tilde{\psi} - u\;b^{z-2a}\;\tilde{\psi}^3
+ b^{a+z}\;\xi.
\end{equation}
Now we define a rescaled noise $\tilde{\xi}=b^{a+z}\xi$, which
has a new intensity
\begin{equation}
\label{eq:rescint}
\tilde{D}=D\;b^{2a+z-d}
\end{equation}
and a correlation time $\tilde{\tau}=\tau/b^z$.
Moreover, in order to keep the shape of the dynamical equation
after the scale transformation, we also define the following
rescaled parameters:
\begin{eqnarray}
\label{eq:rescpar}
\tilde{r}=r\;b^z,\;\;\;\;\;\;\;\;
\tilde{\mu}=\mu\;b^{z-2},\;\;\;\;\;\;\;\;
\tilde{u}=u\;b^{z-2a}
\end{eqnarray}
Now we place ourselves in the critical region of the system
(which in a first approximation is supposed to be located near
$r=0$) where, in the absence of the nonlinear term
($u=0$), the equation remains invariant under
the scale transformation if the exponents $a$ and $z$ are
chosen to be
\begin{eqnarray}
\label{eq:trivexp}
a&=&a_0=\frac{d-2}{2}\;\;\;\;\;\;\;\;\;\;\;\left( \eta_0 = 0 \right)
\nonumber \\
z&=&z_0=2
\end{eqnarray}
These are the {\em trivial exponents} of model
(\ref{eq:tdglm}-\ref{eq:glf}) with
colored noise, which coincide with the ones corresponding to
the white-noise case \cite{hohenberg77}. They are valid when
the nonlinearity of the model does not play a role in the
critical behavior of the system, and this occurs beyond a
critical dimension which we can now determine. As we have seen
in (\ref{eq:rescpar}), the nonlinear coefficient $u$
transforms upon rescaling with an exponent $z-2a$. By making
use of the trivial values of $z$ and $a$, one finds that
this exponent happens to be $z_0-2a_0=4-d\equiv \varepsilon$. Thus when
$d>d_c=4$, the nonlinearity decreases under rescaling
($b>1$), so that $u$ is {\em irrelevant}. On the other hand,
for $d<d_c$, $u$ grows under
rescaling, and hence it is now {\em relevant}. 
This analysis is independent of the character of the noise, i.e.
it does not depend on whether the noise is white or colored.
In fact, since $z>0$ and $\tilde{\tau}=\tau\;b^{-z}$, the
correlation time $\tau$ is always an
irrelevant parameter, so that one should not expect any effect
of it in the universal properties of the model. However it has,
as we will see, influence on the system.

\section{DRG results}
\label{sec:drg-an}
The DRG is going to be implemented in
Fourier space, so the first thing we must do is to transform
our Langevin equation (\ref{eq:tdglm}) conveniently. Let us
define the Fourier transform of a field
variable $\psi$ as:
\begin{eqnarray}
\label{eq:ftdef}
\psi({\bf x},t) = \frac{1}{(2\pi)^{d+1}} \int
\psi({\bf k},\omega) e^{i({\bf k}\cdot {\bf x}- \omega t)}
\;d^dk\; d\omega
\end{eqnarray}
where ${\bf k}$ is a vector in a
d-dimensional (Fourier) space. According to these definitions,
the Fourier transform of Eq. (\ref{eq:tdglm}) can be written as
\begin{eqnarray}
\psi(k,\omega)&=&\psi^{0}(k,\omega) 
\nonumber \\ &-& u\; G_0 (k,\omega)
\int_{k_1\;\omega_1} \int_{k_2\;\omega_2}
\psi({\bf k}_1,\omega_1)\; \psi({\bf k}_2,\omega_2)
\;\psi({\bf k}-{\bf k}_1-{\bf k}_2, \omega-\omega_1-\omega_2)
\label{eq:ft2}
\end{eqnarray}
where $\int_{k\;\omega}$ stands for
$\mathletters{\frac{1}{(2\pi)^{d+1}}
\int d^dk\;d\omega}$.
$\psi^0$ is the zeroth-order aproximation (in $u$)
to the field in momentum space:
\begin{equation}
\psi^{0}({\bf k},\omega) = G_0 (k,\omega)\;
\xi({\bf k},\omega)
\label{eq:psizero}
\end{equation}
and $G_0$ is the propagator:
\begin{equation}
G_0(k,\omega) = \frac{1}{r+\mu k^2-i\omega}
\label{eq:propag}
\end{equation}
On the other hand, by using definition (\ref{eq:ftdef})
in Eq. (\ref{eq:oun}) we can calculate
the correlation of the noise in Fourier space:
\begin{equation}
\label{eq:fourcor}
< \xi({\bf k},\omega)\xi({\bf k}',\omega') > = (2\pi)^{d+1}
\frac{2D}{1+\omega^2\tau^2}\; \delta^d({\bf k}+{\bf k}')
\;\delta(\omega+\omega')
\end{equation}

\vskip8mm
\epsfxsize=5cm
\centerline{\epsfbox{drg1.plt}}
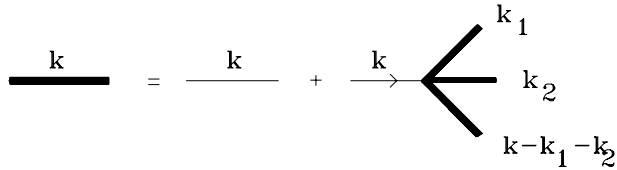
\centerline{\em Diagrammatic version of Eq. (\ref{eq:ft2})}
\centerline{\sc Fig. 1}

DRG will now be applied to Eq. (\ref{eq:ft2}). In order to
simplify this procedure, a diagrammatic notation is very
convenient. The diagram version of Eq. (\ref{eq:ft2}) is shown
in Fig. 1. Thick lines stand for the field,
whereas a thin line represents the zeroth order aproximation
of the field and a vertex stands for the two integrals
(over ${\bf k_1}$ and ${\bf k_2}$) and
$u$. The propagator is represented by a thin line with
an arrow. As can be seen in this figure, momentum through a
vertex must be conserved.

In this work we have decided to
carry out a DRG analysis using the standard momentum shell integration
scheme. While details of this method are well documented in Ref.
\cite{ma76}, here we briefly review the procedure, which consists of the
two following steps:

\paragraph{Momentum Shell Integration.}
The first step of the DRG is to eliminate
the short wavelength modes (i.e. integrate out modes in the momentum shell
$\Lambda e^{-l} < |k| < \Lambda$, where $l > 1$ and $\Lambda$ is the
momemtum upper cutoff), since we are interested in the long wavelength
behavior of the system.

\paragraph{Space Rescaling.}
After the momentum shell intergration step, we
rescale space in such a way that the full momentum space is recovered. This
leads to the differential equations satisfied by the running coupling
constants. A fixed point analysis of these equations will then reveal
the scaling properties of the long wavelength correlations which we
are seeking.

Explicit details of the DRG calculations for our model are summarised
in the Appendix.
The DRG results are obtained from
the analysis of the {\em differential flow equations}
\cite{ma76,fisher83},
defined as the infinitesimal
variation of the parameters when the renormalization step
$l$ is very small. From (\ref{eq:rec}),
they are simply:
\begin{mathletters}
\label{eq:dfe}
\begin{eqnarray}
\label{eq:dfe-r}
\frac{d\bar{r}}{dl}&=&z\bar{r} + 3\;\bar{u}\;K_4 (1-\mu\tau)
\\
\label{eq:dfe-u}
\frac{du}{dl} &=& (z-2a)u-
9 u\bar{u}\;K_4
\\
\label{eq:dfe-mu}
\frac{d\mu}{dl} &=& (z-2)\; \mu
\\
\label{eq:dfe-D}
\frac{dD}{dl}&=& (2a+z-d)\;D
\end{eqnarray}
\end{mathletters}
where the following parameters have been defined:
\begin{eqnarray}
\label{eq:ext-par}
\bar{r} \equiv \frac{r}{\mu},\;\;\;\;\;\;\;\;\;\;\;\;\;\;\;
\bar{u} \equiv \frac{u D}{\mu^2}
\end{eqnarray}

The {\em fixed point} of this transformation is the one at which these
derivatives are zero. Physically it represents the critical
point of the system. Therefore, by studying its position as a
function of $\tau$ we will be able to study the influence of
the time correlation of the noise in the disordering
transition induced by the noise intensity. 
By imposing this invariance on Eqs. (\ref{eq:dfe-mu}) and
(\ref{eq:dfe-D}) one finds that the values of
the exponents $a$ and $z$ up to ${\cal O}(\varepsilon)$ are
the trivial ones already obtained in Sec. \ref{sec:trivial-exp}.
Invariance of the static parameters
(Eqs. (\ref{eq:dfe-r}) and (\ref{eq:dfe-u})) shows that
the fixed point is given by:
\begin{equation}
\label{rec3}
\bar{r}^*=-\frac{\varepsilon}{6}(1-\mu^* \tau),
\;\;\;\;\;\;\;\;\;
\bar{u}^*=\frac{\varepsilon}{9\;K_4}
\end{equation}
For $\tau=0$ (white noise) the critical value of $r$ is negative, which is
reasonable: in the presence of an additive (disordering) noise,
the system will be disordered even for a small negative value of $r$.
When $\tau\neq 0$ the critical value of
$r$ is nearer 0, which means that the disordering effect of the
noise will somehow become diminished because of its correlation in time.
Thus the
effect of the time correlation of the noise is a "softening"
one, as expected and explained above.
It is worth noting that in the white-noise limit our results coincide
with the known values appearing in the
literature \cite{ma76,hohenberg77,fisher83}.

Concerning the evaluation of the critical exponents of the
system, this can be done in a simple way once the recursion
relations are known. By making use of standard techniques
\cite{fisher83} involving the evaluation of the eigenvalues
of the recursion-relation matrix it is easily seen that
there is {\em no} contribution of $\tau$ (up to first order)
to any of the exponents. Hence one can conclude from this
analysis that a correlation in time of the fluctuations
affecting a $\phi^4$ non-conserved model influences the position
of the critical point of the system (as observed numerically
\cite{ojalvo94a}),
but {\em not} its universal properties (through its critical
exponents).

\section{Colored Noise in Space and Time}
\label{sec:space}

A more realistic assumption in relation to external noise is
the existence of a non-delta correlation also in space. Intuition
tells us that the influence of a non-zero correlation length of the
noise will be qualitatively the same as the effect of the correlation
time, i.e. an ordering one. This was checked numerically and also
explained theoretically in Ref. \cite{ojalvo94a}.
A DRG argument concerning this behavior can be made in a
straightforward way following the procedure previously described.
The noise is chosen to be a generalization of Eq. (\ref{eq:oun})
with no delta correlation in space. This can be done in a simple
way by assuming that it obeys the Langevin equation
\begin{equation}
\label{eq:lambda}
\dot{\xi}(\vec{r},t) = -\;\frac{1}{\tau}\;(1-\lambda^2\nabla^2)\; \xi +
\frac{1}{\tau} \;\eta(\vec{r},t)\;,
\end{equation}
where $\eta(\vec{r},t)$ is a gaussian white noise with correlation
(\ref{eq:white}).
This is an extension of the Ornstein-Uhlenbeck process, where
the laplacian term takes into account the coupling of the field
at different points \cite{ojalvo92b}, so that $\lambda$ is the correlation
length. The correlation of this noise in Fourier space can be
seen to be
\begin{equation}
\label{eq:lbdcorr}
< \xi({\bf k},\omega)\xi({\bf k}',\omega') > = (2\pi)^{d+1}
\frac{2D}{\left(1+\lambda^2 k^2 \right)^2+\omega^2\tau^2}\;
\delta^d({\bf k}+{\bf k}')
\;\delta(\omega+\omega'),
\end{equation}
which can be compared to Eq. (\ref{eq:fourcor}).

The scaling analysis of Sec. \ref{sec:trivial-exp} applied to this
new situation shows that $\lambda$ changes as $\tilde{\lambda} =
\lambda / b$, so that it is an irrelevant parameter (as $\tau$).
Nevertheless we aim to analyze, as in the case of the correlation
time, its nonuniversal effects on the system.

The contribution of $\lambda$ to the diagrams which renormalize
$r$ and $u$ can be easily obtained. The new results for the
corresponding differential flow equations are
\begin{mathletters}
\label{eq:dfe-lbd}
\begin{eqnarray}
\label{eq:dfe-r-lbd}
\frac{d\bar{r}}{dl}&=&z\bar{r} + 3\;\bar{u}\;K_4
\;\frac{1}{\left(1+\lambda^2\right)^2}\;
\left(1-\frac{\mu\tau}{1+\lambda^2}\right)
\\
\label{eq:dfe-u-lbd}
\frac{du}{dl} &=& (z-2a)u-
9 u\bar{u}\;K_4\;\frac{1}{\left(1+\lambda^2\right)^2}
\end{eqnarray}
\end{mathletters}
which should be compared to Eqs. (\ref{eq:dfe-r}-\ref{eq:dfe-u}).
These new contributions of $\lambda$ modify the value of the
fixed point in such a way that $\bar{r}^*$ and
$\bar{u}^*$ become: 
\begin{equation}
\label{rec3-lbd}
\bar{r}^*=-\frac{\varepsilon}{6}\left(1-\frac{\mu^* \tau}
{1+\lambda^2}\right),
\;\;\;\;\;\;\;\;\;
\bar{u}^*=\frac{\varepsilon}{9\;K_4}\;\left(1+\lambda^2
\right)^2
\end{equation}
This modification does not lead
to variations of the eigenvalues of the matrix associated to the
transformation, so that critical exponents are not changed from
the white-noise case, as expected.

Hence we can conclude that the noises
used here, either (\ref{eq:fourcor}) or (\ref{eq:lbdcorr}),
with a finite correlation time and length, change the position
of the critical point but not the critical exponents, when compared
to the white-noise case (\ref{eq:white}). This is due to the way
this kind of noises behave (see (\ref{eq:rescint}), for instance).
Should the correlation of the noise
decay as a power law, then, according to the scaling analysis of
Ref. \cite{medina89}, critical exponents would be changed.

\section{Comparison with numerical results}
\label{sec:comp}

In Refs. \cite{ojalvo92a,ojalvo94a} model A (\ref{eq:tdglm})
with noises (\ref{eq:oun}) and (\ref{eq:lbdcorr}) was studied
by means of a numerical simulation in 2-d. Two
main results were obtained there. Firstly, it was established that either
$\tau$ or $\lambda$ stabilize the system, in the sense that the
critical value of the noise intensity is enlarged. Secondly, critical
exponents were also evaluated by means of a finite-size scaling analysis,
and their values were found to be similar (within error bars) to
the accepted values for the white-noise case. Since
the numerical evidence of the critical-noise-intensity shift as
a function of $\tau$ and $\lambda$ is much clearer,
that is what we want to explain
in the light of the above DRG results.

The simulation model, defined by fixed values of $r$, $\mu$ and
$u$, was
\begin{equation}
\label{eq:tdgl}
\frac{\partial\psi(\vec{x},t)}{\partial t} = \frac{1}{2}
\left( \psi - \psi^3 + \nabla^2 \psi \right) + \xi(\vec{x},t)
\end{equation}
where the only independent parameters were
the noise parameters $D$, $\tau$ and $\lambda$. 

Let us now consider Eq. (\ref{eq:tdglm}) at the critical point
with the parameters $r^*$, $u^*$, $D^*$, $\mu^*=1$, $\tau$ and
$\lambda$.
By means of a change of variables, one can recover the simulation
model (\ref{eq:tdgl}) with a critical noise intensity given by
\begin{equation}
\label{eq:cni}
D_c=\frac{1}{2} \;\frac{\bar{u}^*}{\mid \bar{r}^*\mid^{\varepsilon/2}
}=D_c(\tau=0,\lambda=0)\left( \left(1+\lambda^2 \right)^2+ \tau \right)
+ \cal{O}\left(\tau^2\right)
\end{equation}
where $D_c(\tau=0,\lambda=0)=\frac{1}{3 K_4}=\frac{8\pi^2}{3}$.

Certainly one cannot expect a good agreement between a calculation in
4-d extended to 2-d and a numerical simulation of a 2-d model in a
finite discrete lattice. In this sense, the predicted value for 
$D_c(\tau=0,\lambda=0)$ is very different from the numerical result
(0.38). Nevertheless, the ratio $D_c(\tau,\lambda)/
D_c(\tau=0,\lambda=0)$ gives better results, as one can see in Fig. 2.
One can thus conclude that RG calculations have given a reasonable
explanation of the critical results for model A when a colored noise
is considered.

\vskip8mm
\epsfxsize=14cm
\centerline{\epsfbox{drg8.plt}}
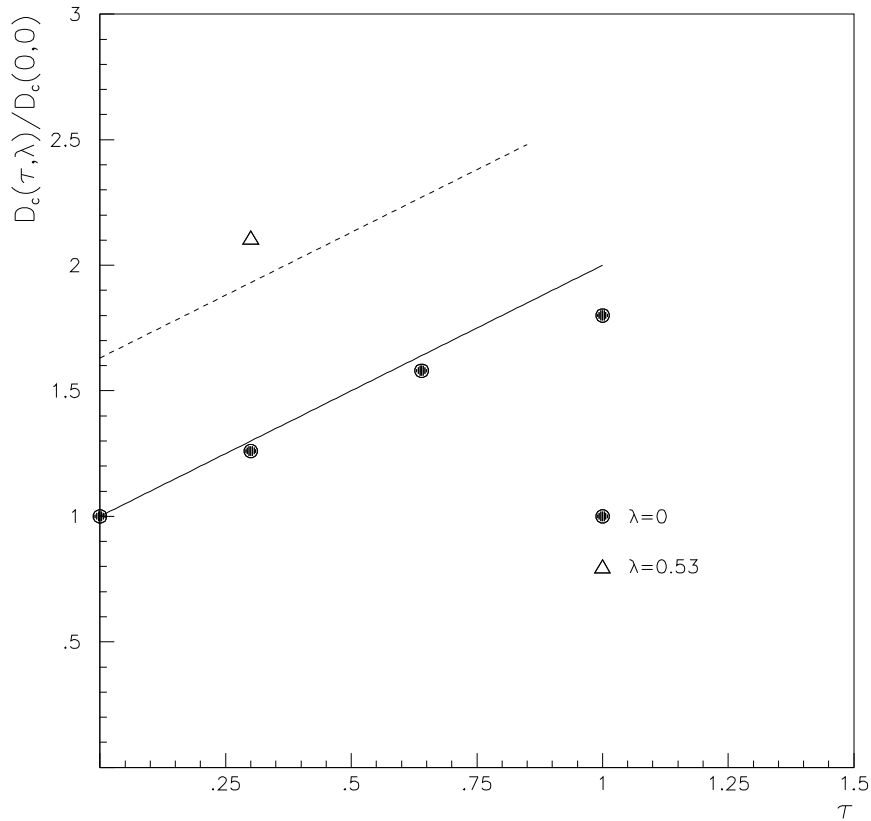
\centerline{\em Ratio of the critical noise intensity in the colored
case to its value in the white}
\centerline{\em case versus $\tau$ (from Ref. \cite{ojalvo94a}).
The lines correspond to the DRG result (\ref{eq:cni}).}
\centerline{\sc Fig. 2}

\begin{acknowledgements}
This research was supported in part by the Direcci\'on General de
Investigaci\'on Cient\'{\i}fica y T\'ecnica (Spain) under Project
No. PB90-0030. H.G. is supported by the Natural Sciences and
Engineering Research Council of Canada, and le Fonds pour la
Formation des Chercheurs et l'Aide \`{a} la Recherche de la
Province du Qu\'ebec.
\end{acknowledgements}

\appendix
\section*{Calculations of the DRG at one-loop order}

In order to eliminate the outer (high-momentum) modes from
Eq. (\ref{eq:ft2})
we break the integral term (the third diagram
in Fig. 1) into modes located in the inner hypersphere
(i.e. those with $0<\mid {\bf k}\mid < \Lambda e^{-l}$),
which will be called {\em external modes}; and modes
located in the outer shell (i.e. those with
$\Lambda e^{-l}<\mid {\bf k}\mid< \Lambda$), which will be
referred to as {\em internal modes}. This leads to a rewriting
of Eq. (\ref{eq:ft2}), whose diagrammatic representation is
shown in Fig. 3 for both the internal and external modes.
Fields and generators depending on internal modes are represented
by means of a slashed line, and those depending on external 
modes are left unchanged.

\vskip5mm
\epsfxsize=8cm
\centerline{\epsfbox{drg2.plt}}
\centerline{(a)}

\vskip5mm
\epsfxsize=8cm
\centerline{\epsfbox{drg3.plt}}
\centerline{(b)}
\centerline{\em Evolution equations for the external (a) and
internal (b) modes}
\centerline{\em ready for the perturbation procedure}
\centerline{\sc Fig. 3}

The evolution equation for the internal field
(Fig. 3b) can be solved
iteratively up to ${\cal O}(u^2)$,
and the result can be introduced in the internal contributions to
the integral term of the corresponding dynamical equation
for the external field (Fig. 3a).
This leads to a renormalized equation where internal modes appear
only in the noise term. After averaging out this internal noise,
the remaining renormalized equation is the one represented in
Fig. 4.

\vskip1cm
\epsfxsize=12cm
\centerline{\epsfbox{drg5.plt}}
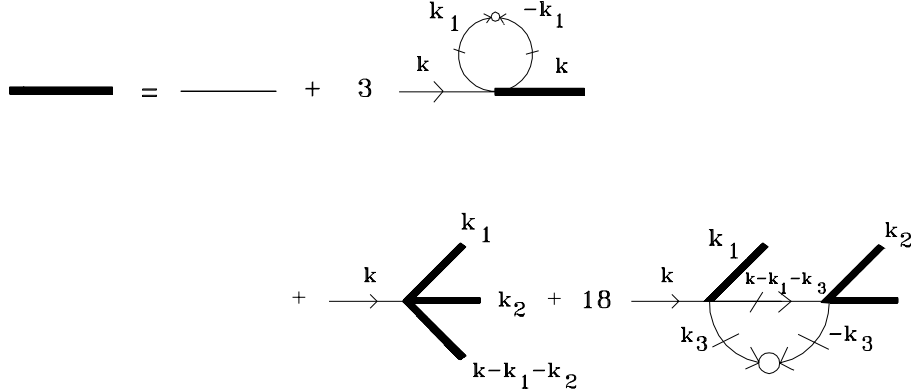
\centerline{\em Renormalized equation after internal noise averaging}
\centerline{\sc Fig. 4}

In order to understand the meaning of the new notation introduced
in Fig. 4, let us make explicit the first diagram.
It comes from an average over the
internal noise term, as shown in Fig. 5. Explicitly,
\begin{eqnarray}
- u\; G_0^{<} (k,\omega)\;
\int_{k_1\;\omega_1}^> \int_{k_2\;\omega_2}^>
\psi^{<}({\bf k}_2,\omega_2)
\left<\psi^{0>}({\bf k}_1,\omega_1)\;\psi^{0>}({\bf k}-
{\bf k}_1-{\bf k}_2,\omega-\omega_1-\omega_2)\right>= \nonumber \\ -
u \;G_0^{<}(k,\omega)\;
\psi^{<}({\bf k},\omega)
\int_{k_1\;\omega_1}^> \frac{2D}{1+\omega_1^2\tau^2}
\;G_0^{>}(k_1,\omega_1)\; G_0^{>}(-k_1,-\omega_1)
\label{eq:avex}
\end{eqnarray}
where definition of $\psi^0$ (\ref{eq:psizero}) and
correlation of the noise in Fourier space (\ref{eq:fourcor})
have been used. We will denote this result as shown in the
second member of Fig. 4. The superindex $<$
($>$) on $G_0$ and $\psi$ denotes an external (internal) mode.

\vskip15mm
\epsfxsize=7cm
\centerline{\epsfbox{drg4.plt}}
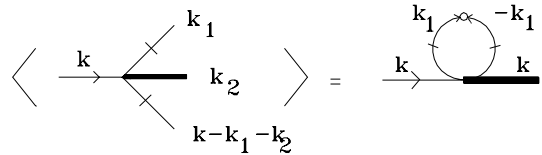
\centerline{\em Interpretation of the diagram which renormalizes $r$}
\centerline{\sc Fig. 5}

Now we will be able to write down the transformation
(recursion) relations for the parameters $r$, $\mu$ and
$u$. We are in the zone where $u$ is relevant
($d<d_c=4$) but are also supposing it to be small (which will
presumably occur for $d \sim 4$). Therefore if these recursion
relations are to be an expansion around $u=0$ they
should also be an expansion in $\varepsilon=4-d$.
Moreover, $r$ will also be supposed to be small in the
following calculations (we are interested in the critical
point).

It should be noted that
the diagram in Fig. 5 contains only one thick line, and
will therefore renormalize the diagram at the left member of
the figure, whereas the second bubble contains three thick
lines, so that it will add to the one-vertex diagram at the
right member of the figure. That means, as we will see, that
the first bubble renormalizes $r$ and the
second one $u$. On the other hand, it is easy to see
that $\mu$ is not renormalized at ${\cal O}(u)$,
since the one-thick-line bubble does not contain any term
proportional to $k^2$.

Concerning the computation of the diagrams, it should be said that
the frequency integrals can be evaluated
by means of contour integrations in the complex plane, and the
integrals in momentum space can also be calculated
by assuming $r\sim 0$ \cite{fisher83} and a very small $\tau$. These integrals
should be expanded in $\varepsilon=4-d$, but
since they are all multiplied by $u$, which is also small, we
only need their zero order in $\varepsilon$. Thus we
evaluate them in $d=4$. We also let $\Lambda=1$.
Moreover, since the parameter $l$ is the step of the
renormalization procedure, and we want this procedure to be
continuous, we will assume $l \ll 1$. After all these
considerations, the final result can be shown to be 
\begin{mathletters}
\label{eq:rr}
\begin{eqnarray}
\label{eq:rr-r}
r_I &=& r + 3 \;\frac{u D}{\mu}\; K_4\left(1-
\mu\tau\right)\;l \\
\label{eq:rr-u}
u_I &=& u - 9 \;\frac{u^2 D}{\mu^2}\; K_4 \;l \\
\label{eq:rr-mu}
\mu_I &=& \mu,
\end{eqnarray}
\end{mathletters}
where $K_d$ is $(2\pi)^{-d}$ times the surface of the unit d-
d-dimensional hypersphere:
\begin{equation}
\label{eq:kd}
K_d\equiv 2^{1-d} \pi^{-d/2} \;\Gamma (d/2)
\end{equation}

We also want to know how the noise intensity D changes under
the effects of the DRG. This is not
given by the analysis explained above, because the parameter D
does not appear explicitly in Eq. (\ref{eq:ft2}). To make it
come out explicitly one can "autocorrelate" this equation, i.e.
multiply the equation for $\psi(k,\omega)$ by the same
equation for $\psi(k',\omega')$. This can be done
diagrammatically in a simple way.
It can be seen that the first diagram renormalizing
$D$ is a two-vertex diagram, which means that it is of ${\cal O}
(u^2)$.
Therefore the recursion relation for $D$ up to the order we are
considering is simply
\begin{equation}
\label{eq:rr-D}
D_I=D
\end{equation}
It is worth noting that there is no first-order correction
in $\tau$ neither for $\mu$, $u$ nor $D$.

Finally, we want the DRG procedure to lead to a renormalized equation
as similar as possible to the original one. At this moment
both equations are identical in form, the only difference
being the range of the variable in momentum space to which
they are applied. Indeed, the renormalized equation only
governs the evolution of the external modes. In
order to eliminate this difference, we shall perform the
following rescalation:
\begin{eqnarray}
\label{eq:resc1}
\tilde{k}=e^{l}k,\;\;\;\;\;\;\;\;\;
\tilde{\omega}=e^{zl}\omega, \;\;\;\;\;\;\;\;\; 
\tilde{\psi}_r=e^{al}\psi_r
\end{eqnarray}
This scale transformation is identical to the one described in
Sec. \ref{sec:trivial-exp} with $b=e^l$. There we could see
that the rescalation led to effective values of
the parameters of the model ((\ref{eq:rescint}) and (\ref{eq:rescpar})).
Hence, by combining the momentum-shell integration process
and the space rescalation, one finds the following final
discrete recursion relations for the model parameters up to
first order in $l$:
\begin{mathletters}
\label{eq:rec}
\begin{equation}
\label{eq:rec-r}
\tilde{r}=r_I\; e^{zl} \simeq r+l\left[z\;r + 3\;K_4
\;\frac{u D}{\mu} \left( 1-\mu \tau
\right) \right]
\end{equation}
\begin{equation}
\label{eq:rec-u}
\tilde{u}=u_I\; e^{l(z-2a)} \simeq u + l
\left[(z-2a)u - 9\;K_4\; \frac{u^2 D}{\mu^2}
\right]
\end{equation}
\begin{equation}
\label{eq:rec-mu}
\tilde{\mu}=\mu_I\; e^{l(z-2)} \simeq \mu + l (z-2)\mu
\end{equation}
\begin{equation}
\label{eq:rec-D}
\tilde{D}= D_I \;e^{l(2a+z-d)} \simeq D + l (2a+z-d) D
\end{equation}
\end{mathletters}

\end{document}